\documentclass[12pt,amssymb,amsfonts]{article} 
\usepackage{latexsym}

\newcommand {\be}{\begin{equation}}
\newcommand {\ee}{\end{equation}}
\newcommand{\bea}{\begin{eqnarray}}
\newcommand{\eea}{\end{eqnarray}}
\newcommand{\ba}{\begin{array}}
\newcommand{\ea}{\end{array}}
\newcommand{\beq}{\begin{eqnarray*}}
\newcommand{\eeq}{\end{eqnarray*}}

\newcommand{\ds}{\displaystyle}

\renewcommand{\Re}{{\rm Re}}
\renewcommand{\Im}{{\rm Im}}

\newcommand{\W}{{\cal W}}
\renewcommand{\k}{{\bf k}}
\newcommand{\x}{{\bf x}}

\begin{document}

\begin{center}
{\Large\bf  Quotient Equations and Integrals of Motion for
Scalar Field}

\bigskip

S.S. Moskaliuk

\medskip

Bogolyubov Institute for Theoretical Physics\\
Metrolohichna Str., 14-b, Kyiv-143, Ukraine, UA-03143\\
e-mail: mss@bitp.kiev.ua
\end{center}

\bigskip

\begin{center}
{\bf Abstract}
\end{center}

\medskip

In this article a group-theoretical aspect of the method of dimensional
reduction is presented. Then, on the base of symmetry analysis for an
anisotropic space geometrical description of dimensional reduction of
equation for scalar field is given. Formula for calculating components
of the energy-momentum tensor from the variables of the field
factor-equations is derived.

\newpage

Much well-known researches [1,2] are devoted to quantization of scalar
field in various anisotropic models. Here we shall briefly outline the
results of this research as applied to our task, following [2].

The covariant equation for massive scalar conform-invariant field in
metric

\be
ds^2=dt^2-\sum_{i=1}^3 A_i^2(t) (dx^i)^2
\ee
is of the form
\beq
\nabla_\mu\nabla^\mu\varphi+(m^2+R/6)\varphi=0,
\eeq
where $\nabla_\mu$ is covariant derivative and $R$ is scalar curvature.

The solution of this equation can be written as
$\varphi_k(x)=(16\pi)^{-3/2} a(\eta)^{-1} \times \nonumber \\
\!\!\!\times \nonumber  \tilde{g}_k(\eta) 
\exp(-i\k\x)$ [3,4].  The temporal function $\tilde{g}_k(\eta)$ 
satisfies the equation of oscillator type

\be
\ddot{\widetilde{g}_k}+ [\mu^2(\eta)K_0^2(\eta) +Q(\eta)
\tilde{g}_k=0.
\ee
Here and after in this article we shall use the following notation:

\bea
k &\!\!\!=\!\!\!& (K_0,\k),\nonumber \\
\k &\!\!\!=\!\!\!&
(k_1,k_2,k_3)=k\left( \sin(\theta)\, \cos(\varphi),\, \sin(\theta)\,
\sin(\varphi), \,\cos(\theta)\right), \nonumber \\
K_0^2(\eta,\k) &\!\!\!=\!\!\!&
k^2+m^2g^2(\eta,\theta,\varphi), \nonumber \\
g &\!\!\!=\!\!\!&  a(\eta)/\mu(\eta,\theta,\varphi), \quad
C_i=\dot{\alpha}_i/\alpha_i, \nonumber \\
\mu^2(\eta,\theta,\varphi) &\!\!\!=\!\!\!&
{\sin^2(\theta)\, \cos^2(\varphi)\over \alpha_1^2(\eta)}
+{\sin^2(\theta)\, \sin^2(\varphi)\over \alpha_2^2(\eta)} 
+{\cos^2\varphi\over \alpha_3^2(\eta)}, \nonumber \\ Q 
&\!\!\!=\!\!\!& \left[(C_1-C_2)^2 +(C_2-C_3)^2 +(C_1-C_3)^2\right].  
\eea 
Following Lagrange method, we pursue the solutions of the 
equation (2) in the form

\beq
\tilde{g}_k(\eta)  &\!\!\!=\!\!\!&
(\mu K_0)^{-1} \left[ \stackrel{*}{\ds\alpha}_k e_+
+\beta_k e_-\right],\\
\dot{\tilde{g}}(\eta)  &\!\!\!=\!\!\!&
-i\mu K_0\left[ \stackrel{*}{\ds\alpha_k} e_+ -\beta_ke_-\right],\\
e_\pm  &\!\!\!=\!\!\!& \exp\left(\pm i
\int\limits_{{\ds\eta}_0} K_0(\eta') d\eta'\right),
\eeq
rendering instead of one differential equation of the second order for
$\tilde{g}_k(\eta)$  a set of two linear equations of the first order
for complex-valued functions $\alpha_k(\eta)$ and $\beta_k(\eta)$,
related by additional condition
$|\alpha_k(\eta)|^2-|\beta_k(\eta)|^2=1$:

\bea
\dot{\alpha}_k &\!\!\!=\!\!\!& \left( {\W\over2} -i
{\widetilde{\W}\over2}\right) \stackrel{*}{\ds\beta}_ke^2_+ -i
{\widetilde{\W}\over2} \alpha_k,\nonumber \\
\stackrel{*}{\ds\dot{\beta}}_k &\!\!\!=\!\!\!&
\left( {\W\over2} +i{\widetilde{\W}\over2}\right) \alpha_ke_-^2 +i
{\widetilde{\W}\over2}\stackrel{*}{\ds\beta},
\eea
where $\W=\dot{\mu}/\mu+\dot{K}_0/K_0$; $\widetilde{\W}=Q/(\mu K_0)$.

In actual practice it is more convenient to change from two
complex-valued functions $\alpha_k(\eta)$ and $\beta_k(\eta)$ to three
real-valued functions

\bea
S_k(\eta) &\!\!\!=\!\!\!& |\beta_k|^2,\nonumber \\
U_k(\eta) &\!\!\!=\!\!\!& 2\Re (\alpha_k\stackrel{*}{\ds\beta}_k
e_-^2),\nonumber \\
V_k(\eta)  &\!\!\!=\!\!\!& 2\Im (\alpha_k
\stackrel{*}{\ds\beta}_k e_-^2),
\eea
for which a set of three linear equations can be derived [5]

\bea
\dot{S}_k &\!\!\!=\!\!\!&  {\W\over2}U_k+{\widetilde{\W}\over2}V_k,
\nonumber \\
\dot{U}_k &\!\!\!=\!\!\!&  \W(2S_k+1) -(\widetilde{\W}+2K_0)V_k,
\nonumber \\
\dot{V}_k &\!\!\!=\!\!\!&  \widetilde{\W}(2S_k+1) 
-(\widetilde{\W}+2K_0)U_k 
\eea 
with initial conditions  
$S_k=U_k=V_k=0$ at $\eta=\eta_0$.

As it can be seen, the structure of  the equations (6), is similar to
that of the set of equations, derived for vector massless files and
massive spinor field [6,7].

The introduced functions and notation enable us to write vacuum
averages for normally ordered operator of energy-momentum tensor
$T_{\mu\nu} = \langle 0_{in}|N_\eta T_{\mu\nu}(\eta,\x)|0_{in}\rangle$
for scalar field [2]:

\bea
T_0^0(\eta) &\!\!\!=\!\!\!&
{1\over(2\pi)^3a^4(\eta)} \int d^3 \k \mu(\eta,\theta,\varphi) 
K_0(\eta,\k) \left(S_k(\eta)- \right. \nonumber \\ 
&\!\!\!-\!\!\!& 
 \left. {Q(\eta)\over 2\mu^2(\eta,\theta,\varphi) K_0^2(\eta,k)} 
\left[ S_k(\eta)+{1\over2} U_k(\eta)\right]\right), \nonumber 
\\[0.2cm] 
T_i^i(\eta)  &\!\!\!=\!\!\!& {1\over(2\pi)^3a^4(\eta)} \int 
d^3\k \mu(\eta,\theta,\varphi) K_0(\eta,\k) \times \nonumber \\ 
 &\!\!\!\times\!\!\!& \left( {k_i^2\over \alpha_i^2\mu^2K_0^2} \left[
S_k(\eta) +{1\over2} U_k(\eta) \right] -{1\over6} U_k(\eta)+
\right.\nonumber \\
 &\!\!\!+\!\!\!&  \left. {1\over6\mu^2K_0^2} (C_i-Q)
\left[ S_k(\eta)+{1\over2} U_k(\eta)\right] -{C_i\over18\mu K_0} V_k
\right), \nonumber \\[0.2cm]
T_\mu^\mu(\eta)  &\!\!\!=\!\!\!&
{1\over(2\pi)^3a^4 (\eta)} \int d^3 
\k \mu(\eta,\theta,\varphi)K_0(\eta,\k) \times \nonumber \\ 
 &\!\!\!\times\!\!\!& \left( {m^2g^2(\eta,\theta,\varphi)\over
K_0^2(\eta,\k)} \left[ S_k(\eta) +{1\over2} U_k(\eta) \right]\right).
\eea

In such a manner, solving the set of equations (6) and examining the
properties of functions $S_k$, $U_k$ and $V_k$, we get the information
on energy-momentum tensor for scalar field.

The author is grateful to Yu.A. Danilov for helpful discussions.

\end{document}